\documentclass{aa}  

\usepackage{graphicx}
\usepackage{txfonts}
\usepackage{graphicx}
\usepackage{epsf,amsfonts,amsmath,amssymb}
\usepackage{mathrsfs}
\begin{document}

   \title{What can we learn from ``internal plateaus"? \\The peculiar afterglow of GRB 070110}

   \author{P. Beniamini \inst{\ref{inst1}},\inst{\ref{inst2}} \& R. Mochkovitch \inst{\ref{inst1}} }

  \authorrunning{P. Beniamini \& R. Mochkovitch}
  \titlerunning{The peculiar early afterglow of GRB 070110}
  
  \institute{UPMC-CNRS, UMR7095, Institut d'Astrophysique de Paris, F-75014 Paris, France \label{inst1} \and Department of Physics, The George Washington University, Washington, DC 20052, USA \label{inst2}}

 
  \abstract
   {The origin of GRBs' prompt emission is highly debated. Proposed scenarios involve various dissipation processes above or below the photosphere of an ultra-relativistic outflow.}
   {We search for observational features that would favour one scenario over the others by constraining the dissipation radius, the outflow magnetization or by indicating the presence of shocks. Bursts showing peculiar behaviours can emphasize the role of a specific physical ingredient, which becomes more apparent under certain circumstances.}
   {We study GRB 070110, which exhibited several remarkable features during its early afterglow: a very flat plateau terminated by an extremely steep drop and immediately followed by a bump. We model the plateau as photospheric emission from a long lasting outflow of moderate Lorentz factor ($\Gamma\sim 20$) which lags behind an ultra-relativistic ($\Gamma> 100$) ejecta responsible for the prompt emission. We compute the dissipation of energy in the forward and reverse shocks resulting from this ejecta's deceleration by the external medium.}
   {Photospheric emission from the long-lasting outflow can account for the plateau properties (luminosity and spectrum) assuming some dissipation takes place in the flow. 
   	The geometrical timescale at the photospheric radius is so short that the observed
   	decline at the end of the plateau likely corresponds to the actual shut-down of the central engine's activity. The following bump results from dissipated power in the reverse shock, which develops when the slower material catches up with the initially fast component, after it had been decelerated.}
   {The proposed interpretation suggests that the prompt phase resulted from dissipation above the photosphere while the plateau had a photospheric origin. If the bump is produced by the reverse shock, it implies an upper limit ($\sigma \lesssim 0.1$) on the magnetization of the slower material. }


   \maketitle
%

\section{Introduction}
The {\it Swift} and {\it Fermi} satellites, launched in 2004 and 2008 respectively,
have dramatically improved our knowledge 
of the observational properties of Gamma Ray Bursts (GRBs). The rapid slewing capabilities of {\it
	Swift} and its narrow field X-ray and visible instruments
XRT and UVOT have revealed the rich phenomenology of the early afterglow and allowed a redshift
determination for nearly 500 bursts. The broad spectral
coverage with {\it Fermi} GBM and LAT from a few keV to the GeV range has shown in some cases the presence of
additional components to the simple Band spectrum
such as power-law spectra extending to very high energies \citep{Swenson2010,Guetta2011,Beniamini2011,Ackermann2013} or possible thermal emission from
the photosphere \citep{Guiriec2011,Axelsson2012}.

On the theoretical side, the dissipation process responsible for the prompt emission
remains uncertain. There are three main
contenders: internal shocks \citep{ReesMeszaros1994,Kobayashi1997,DM1998,DM2000,Beloborodov2000,Bosnjak2009}, magnetic reconnection \citep{Usov1994,DS2002,Giannios2008,Zhang2011,McKinney2012,Sironi2015,BG2016,Granot2016,Kagan2016,BG2017} or dissipative photosphere \citep{Thompson1994,Ghisellini1999,Meszaros2000,Rees2005,Peer2005,Giannios2007,Lazzati2010,Beloborodov2010,Levinson2012,Beloborodov2013},
which differ in terms of emission radii
and radiative mechanisms. Internal shocks and magnetic reconnection 
respectively dissipate kinetic and magnetic energy in the flow, which is then radiated 
above the photosphere, typically via the synchrotron process \citep{Katz1994,ReesMeszaros1994,Sari1996,Kumar2008,Daigne2011,Beniamini2013,Beniamini2014}. In photospheric models
dissipation below the photosphere produces energetic electrons that can transform a seed 
Planck spectrum into a broken power-law with Inverse Compton (IC) collisions contributing to the high-energy emission and an additional
synchrotron component at low energies \citep{Thompson1994,Meszaros2000,Giannios2006,Beloborodov2010,Lazzati2010,Vurm2011,Giannios2012}.

Various proposals have been made to explain the unexpected features of the early afterglow (extended plateaus, flares, steep breaks) often involving a long term activity of the burst central engine or alternatively the contribution of a long-lived reverse shock \citep{Uhm2007,Genet2007}.
This observational diversity can be seen as a source of confusion and in some
respects, it certainly is. However, peculiarities could also result from one or several ingredients of the model taking non standard values.
In this case, they can in fact provide clues to elucidate the physics at work.
We concentrate here on GRB 070110 and use it as a laboratory to constrain the dissipation 
process, the importance of photospheric emission and the role played by the reverse shock.

GRB 070110 started with a rather regular prompt emission (see \S \ref{sec:obs}). 
{\it Swift} XRT observations followed at 100 s 
and showed the usual early steep decay with the X-ray flux $F_{\rm X}(t)\propto
t^{-2.53}$. After a data interruption due to orbit constraints,
the XRT light curve exhibited a very flat plateau from a few $10^3$ s to 18 000 s
(300 -- 5400 s in the burst rest frame at $z=2.35$). This 
plateau ended with a very steep drop ($F_{\rm X}(t)\propto t^{-9}$). Remarkably, the plateau is not
seen in simultaneous observations 
with the UVOT instrument, which shows a regular power-law behaviour.

The initial decay following the prompt phase can be explained as high-latitude emission from the final emitting shell at $R\sim 2\,c\Gamma^2 \tau$ where $\tau$ 
is the duration of the prompt activity. This value for the radius is naturally expected
in internal shocks models \citep[e.g.,][]{Kobayashi1997,Hascoet2012} and in some reconnection models \citep[e.g.,][]{BG2016}. Conversely, in photospheric models,
where the emission radius is much smaller, this geometrical interpretation
does not work and the decay should correspond to an effective decline of the
central engine's activity. In some respects, the situation at the end of the plateau is just the opposite:
the drop is much too steep to be related to either a forward or reverse shock contribution.
This has led to the idea of ``internal plateaus'' \citep{Zhang2006,Liang2006,Troja2007}, where the dissipation
radius is much closer to the central source.
The geometrical time-scale $R/2c\Gamma^2$ can then be very short but the physical time-scale for the decay 
of source activity must also be short, since the observed decline is controlled by the longest of the two.

One possible way to obtain a short physical decay time-scale is to suppose that the
central source that powers the plateau is a magnetar that 
eventually collapses to a black hole. However, a short decay time-scale requires also a short geometrical time-scale.
The latter could be naturally produced if the plateau results from photospheric emission.
We discuss below these two possibilities. After a short summary of the
observational properties of GRB 070110 in \S \ref{sec:obs} we consider the magnetar hypothesis
in \S \ref{sec:magnetar}. \S \ref{sec:photo} compares the photospheric emission of a steady outflow to the
plateau data while \S \ref{sec:RevFor} presents the forward and reverse shock contributions to
the afterglow. Our results are discussed in \S \ref{sec:summary}, which is also the conclusion.

\section{GRB 070110: brief summary of the observational properties}
\label{sec:obs}
The prompt phase of GRB 070110 was not especially remarkable. It lasted 90 s
($t_{90}$ [15-150 keV]=89 $\pm 7$ s) with an initial peak followed by 4-5
overlapping pulses during the first 40 s
and a tail for the next 50 s. The time-averaged photon spectrum could be
fitted by a single power law of index 1.57, which
did not allow to constrain the peak energy. The isotropic energy released
in the BAT spectral range 15-150 keV (50-500 keV
in burst rest frame) was $E_{\rm iso, BAT}=2.3\,10^{52}$ erg. Assuming
that the spectrum keeps the same slope from 0 to the peak energy
$E_{\rm p}\ge 500$ keV and then breaks to a photon index $\beta>2$, the
total isotropic energy would be
\begin{equation}
E_{\gamma,\rm iso}\approx \left(3.8+{1.6\over \beta-2}\right)\left({E_{\rm
		p}\over 500\ {\rm keV}}\right)^{0.43}\times 10^{52}\ \ \ {\rm erg}\ .
\end{equation}
The flux at the end of the prompt phase decays as $t^{-2.45}$, which is
typical of the evolution observed in most GRBs. The
plateau phase that follows is very flat and extends to 20 ks (or
$\tau_{\rm plateau}=5800$ s in the rest frame) with a luminosity $L_{\rm
	plateau}\sim 10^{48}$
erg.s$^{-1}$ within the XRT spectral range, which covers 1.5 decades in
energy. This value of the luminosity therefore corresponds
to a lower limit. If the spectrum keeps the same slope ($F(E)\sim E^{-1}$)
in an interval [$E_1,E_2$] covering $d>1.5$ decades
the corrected luminosity will be $L_{\rm plateau}\sim 10^{48}\,(d/1.5)$
erg.s$^{-1}$, leading to a total energy radiated during the plateau phase
$E_{\rm plateau}=L_{\rm plateau}\times \tau_{\rm
	plateau}=5.8\,10^{51}\,(d/1.5)$ erg.

Since the plateau
is not seen at optical wavelengths the observed spectrum should break
at some energy below the lower bound of the XRT range of 0.3 keV. This
implies an average slope $s<0.4$ (with $F(E)\propto E^{-s}$)
in the interval from 0.3 keV to the optical. The plateau abruptly ends
with the X-ray flux decreasing as $t^{-9}$, or on a (observed)
time scale of about 5000 s, corresponding to $\delta\tau_{\rm
	plateau}/\tau_{\rm plateau}\sim 0.25$.
This implies that the plateau emission cannot come from either the forward
or reverse shocks, which would imply
$\delta\tau_{\rm plateau}/\tau_{\rm plateau}\sim 1$.

Just following the drop, a bump is observed in the XRT light curve with an
increase of the flux by a factor of 3 and a total
duration of about $10^5$ s \citep{Troja2007}. Data is lacking in the optical so that it is
not possible to know if the bump would have been also visible in
the optical light curve. A possible second weaker bump is also seen
immediately after the first one before the X-ray light curve recovers
a power law decline of temporal index $\alpha\approx -0.6$.
As noted by \cite{Troja2007}, this decay does not follow the regular
closure relations as expected in the forward afterglow scenario, given the spectral index measured
in the X-ray band at the same time (approximately $F_{\nu}\propto \nu^{-1.1}$).

The most striking feature in the afterglow of GRB 070110 is clearly the
very steep drop at the end of the plateau, which contrasts with
the standard decay at the end of the prompt phase. This internal plateau
must be directly related to the activity of the central engine,
showing that at least in some bursts it can still operate (at a reduced
level) for a duration that can exceed by more than two
orders of magnitude that of the prompt phase.

\section{Origin of the plateau: the magnetar hypothesis}
\label{sec:magnetar}
Several studies have suggested a spinning down highly magnetized pulsar (``magnetar") as the origin for the plateau of GRB 070110 \citep{Troja2007,Lyons2010,Yu2010,Du2016}. The magnetar solution must satisfy two main constraints: ({\it i}) the available energy reservoir should be 
able to power the plateau for 1.5 hours and ({\it ii}) the subsequent drop of luminosity should be steep enough. For a dipole spin-down mechanism, the released power at $t>t_s$ (where $t_s$ is the starting time of the spin-down) can be written as:
\begin{equation}
\label{eq:magnetar}
L=L_0 (1+(t-t_s)/t_0)^{-2}
\end{equation}
where $t_0=3c^3 I(1+z)/B_p^2 R^6 \Omega_0^2$ is the observed spin-down time of the pulsar and depends on the dipolar magnetic field strength at the poles ($B_p$), the pulsar's moment of inertia ($I$), its radius ($R$) and initial spin frequency ($\Omega_0$). The initial power is 
then given by the ratio of the initial rotational energy to the spin-down time: $L_0\approx(1/2)I\Omega_0^2/t_0$. One therefore expects a constant power $L_0$ up to $t_0$ followed by a power-law decline.
If this power is directly translated to the observed X-ray plateau, it is possible to link the observed luminosity and duration to the initial magnetar parameters, i.e. $B_p$ and $\Omega_0$. Indeed, applying this method to GRB 070110 (assuming $I=10^{45}\mbox{g cm}^2, R=10^6$cm), \cite{Troja2007} find 
$B_p \gtrsim 3 \times 10^{14}$ Gauss and $P_0\equiv 2 \pi / \Omega_0 \lesssim 1$msec. 
The magnetar should therefore be spinning very close to the break-up limit \citep{Lattimer2004} with $P_0=0.96$msec.
This implies that unless almost all of the spin-down luminosity is translated to the observed X-ray emission, the energetic requirements may be too large 
to be accounted for by a magnetar.

Note however that this calculation has two important caveats. First, it assumes 100$\%$ efficiency in translating the magnetar's luminosity to the observed one in X-rays. Furthermore, it implicitly assumes that the flux is rather sharply peaked in the X-ray band, possibly suggesting a thermal spectrum and photospheric origin for this emission (see further discussion of the spectrum in \S \ref{sec:dissipation}). Secondly, it assumes that the luminosity is released isotropically. Breaking the first of these assumptions would require even more energy at the source while the latter reduces the energy requirements. A full assessment of the magnetar parameters in this model (and its viability from an energetic point of view) therefore crucially depends on the efficiency of energy conversion and on the opening angle of the jet.
Regardless of the implied magnetar parameters, a preliminary question that has to be addressed in this scenario is: {\it what accounts for the prompt emission?}
If the magnetar spin-down luminosity is associated with that of the plateau then it seems unlikely that it could be also associated with the prompt emission (which involved different luminosities and time-scales, see \S \ref{sec:obs}) (see also \citealt{Granot2015}).

As pointed out by \cite{Troja2007}, the rapid decay at the end of the plateau (at $t_{\rm obs}\sim 2\times 10^4$ s) suggests an ``internal" origin for the plateau emission, 
i.e. the magnetar's power should be dissipated at a radius well below that of the forward shock, indicating that the emission cannot be simply 
due to energy injection \citep[e.g.][]{ZhangMeszaros2001}. This is because the observed decay is on a time-scale $t_{\rm decay}< 5\times 10^3$ s, significantly shorter than the angular time-scale associated with the forward shock material at that time: 
$t_{\theta}=R(1+z)/2c\Gamma^2\sim t_{\rm obs}\sim 2\times 10^4$ s.
Also, the magnetar's energy cannot be emitted directly from its surface as the magnetar is highly super-Eddington. This leads to a lower limit on the radius and Lorentz factor at the dissipation radius.
Therefore, in the context of the magnetar model, one must still find some alternative way of dissipating the power as well as explaining why the emission should reside predominantly in X-rays.

\begin{figure}
	\includegraphics[width=0.45\textwidth]{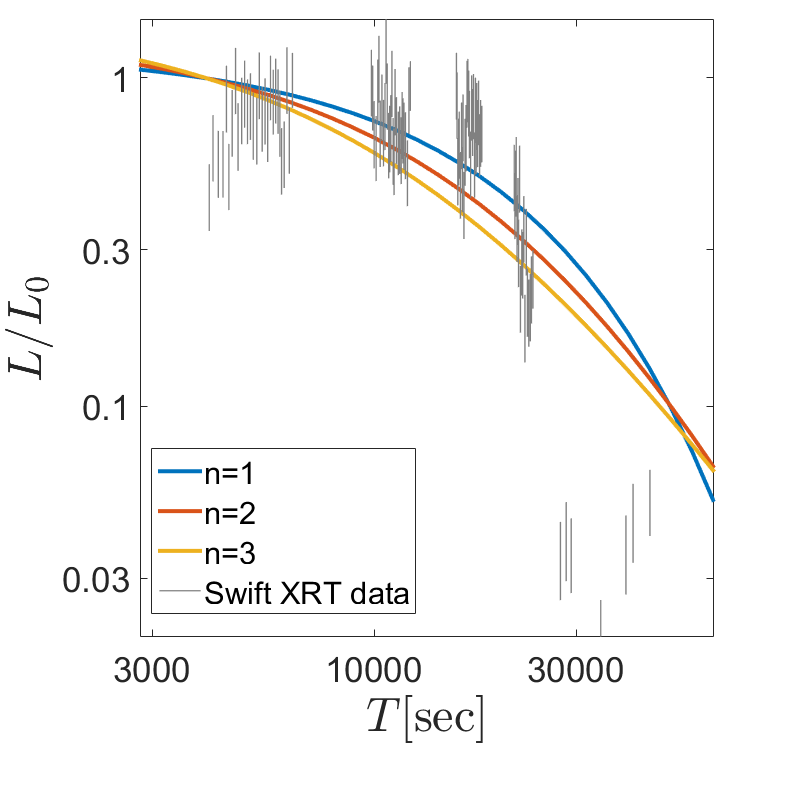}
	\caption
	{\small Normalized luminosity ($L/L_0$) as a function of time (see Eq. \ref{eq:magnetargen}) for various spin-down models ($n=1,2,3$) as compared with the observed XRT data. We take $t_0=2\times 10^4$ s which is the observed time of the end of the plateau and $t_s=4000$ s, which is the observed time at which the plateau starts and is an upper limit on the actual value. However, the results are largely insensitive to this choice. Clearly, spin-down models fall too slowly as compared with the drop at the end of the plateau.}
	\label{magnetarfig}
\end{figure}

An additional concern for the magnetar model has to do with the shape of the decline at the end of the plateau. Namely, the decline expected from a dipole spin-down (Eq. \ref{eq:magnetar}) is too shallow as compared with observations. For a general spin-down mechanism, one can write:
\begin{equation}
\label{eq:magnetargen}
L\!=\!
\left\{ \!
\begin{array}{l}
L_0 (1+\frac{n-1}{2}\frac{t-t_s}{t_0})^{1+n\over 1-n} \quad n \neq 1\\
\\
L_0 e^{-\frac{t-t_s}{t_0}} \quad n=1\\
\end{array} \right.
\end{equation}
where the expression in Eq. \ref{eq:magnetar} is restored for $n=3$.
In Fig. \ref{magnetarfig} we plot this function for a range of values of $n=1-3$ ($n=1-2.8$ is the typically inferred range from observations of isolated pulsars, e.g. \citealt{Hamil2015}) and compare with the observed light-curve.
Generally, the slope becomes steeper as $n$ decreases, but even for $n=1$, it is clear that the drop is too shallow as compared with observations. 
An alternate possibility, is that the spin-down stops abruptly due to the collapse of the rotating neutron star to a black hole. This could produce an almost step-like drop in the light-curve, but may require fine-tuning in order for the collapse to happen at $t_{\rm coll}\approx t_0$ or else, if $t_{\rm coll}\ll t_0$, would require even larger values for the magnetar's initial energy which is already uncomfortably close to the maximum theoretical value.

\section{Photospheric emission from a relativistic outflow}
\label{sec:photo}
\subsection{Basic equations: non dissipative outflow from the central engine to the photosphere}
\label{sec:nodiss}
Considering the various potential issues with the magnetar hypothesis outlined above, we suggest a simple alternative where the central engine is instead powered by a black hole.
In this scenario the central engine transitions from an initial phase with high injected power 
driving a jet of large Lorentz factor (several hundreds) responsible for the prompt emission to a second phase with much lower injected power driving a slower outflow 
with a Lorentz factor of a few tens only. \footnote{One may tentatively propose that these two phases are respectively powered by the extraction of the black-hole 
	rotational energy via the Blandford-Znajek mechanism and by the release of gravitational energy from  
	accretion.}       
The plateau 
corresponds to the photospheric emission from this slower outflow.   
We suppose first that the outflow is non-dissipative and starts with $\Gamma_0\approx 1$ from an initial radius $R_0$, close to the central engine (i.e. $R_0\lesssim 10^7$ cm). 
It then accelerates and expands adiabatically up to the photosphere. The initial temperature $T_0$ at the origin can be estimated by
\begin{equation}
{\dot E}_{\rm th}=\epsilon_{\rm th}{\dot E}=4\pi R_0^2\times a T_0^4\,c\ ,
\end{equation}
where $ \dot E$ is the total power injected in the flow, $\epsilon_{\rm th}$ is the fraction of that energy in thermal form and $a$ is the radiation constant. This leads to
\begin{equation}
T_0=\left({\epsilon_{\rm th} \dot E\over 4\pi\,ac\,R_0^2}\right)^{1/4}=0.36\,\, \epsilon_{\rm th}^{1/4}R_{0,7}^{-1/2}{\dot E}_{50}^{1/4}\ \ {\rm MeV}\ 
\end{equation}
where here and elsewhere in the text we use the notation $q_x$ for $q$ in units of $10^x$ in cgs.
Since already at the base of the jet, the temperature is below the pair creation threshold (and at larger radii the temperature only further decreases), we can safely ignore the effects of pair creation within the flow.
In addition, in order for the photospheric emission from this material to be able to account for the X-ray plateau, we can also assume that the flow has stopped accelerating by the time it reaches the photosphere. This is because, otherwise, assuming a pure fireball in which the acceleration persists as $\Gamma \propto r$, the observed peak of the photospheric component would have been $\approx 3k_B T_0/(1+z)\approx 300$keV \citep[e.g.,][]{KumarZhang2015}, and the contribution of this emission at the observed [0.1,10]keV X-ray band would have been too small \footnote{In fact, even if the acceleration is more gradual, for example $\Gamma \propto r^{1/3}$ as expected for MHD jets \citep{Drenkhahn2002}, the observed temperature when acceleration is incomplete at the photosphere is still expected to be $\sim 100$keV with a relatively weak dependence on the jet parameters \citep{GianniosSpruit2005}. This result can also be extended to generic acceleration models as shown by \cite{Giannios2012}}.
With these assumptions, the luminosity at the photosphere takes the form 
\begin{equation}
\label{eq:Lph}
L_{\rm ph}=\epsilon_{\rm th}{\dot E}\,\left({R_{\rm ph}\over R_0}\right)^{-2/3}\Gamma^{2/3}\ ,
\end{equation}
where $R_{\rm ph}$ is the photophereric radius and $\Gamma$ is the Lorentz factor at $R_{\rm ph}$.
Similarly, the observed temperature is given by
\begin{equation}
\label{eq:Tph}
T_{\rm obs}={T_0\over 1+z}\,\left({R_{\rm ph}\over R_0}\right)^{-2/3}\Gamma^{2/3}\ .
\end{equation}

The photospheric radius can be expressed by (see also \citealt{Zhang2011})
\begin{equation}
\label{eq:Rph}
R_{\rm ph}\approx {\sigma_{\rm T}\dot E \over 8\pi m_{\rm p} c^3 \Gamma^3(1+\sigma)}\approx {6\,10^{13}\over (1+\sigma)}\,{{\dot E}_{50}\over \Gamma_1^3}\ \ {\rm cm}
\end{equation}
where $\sigma_{\rm T}$ is the Thomson cross section, $m_{\rm p}$ the proton mass
and $\sigma$ the magnetization of the flow such that ${\dot E}_{\rm K}={\dot E}/(1+\sigma)$ is
the injected kinetic power. Using Eqs.(\ref{eq:Lph}), (\ref{eq:Tph}) and (\ref{eq:Rph}) we obtain
\begin{equation}
\label{eq:Lph2}
L_{\rm ph}=1.4\,10^{46}\,\epsilon_{\rm th}\left(1+\sigma\right)^{2/3}R_{0,7}^{2/3}{\dot E}_{50}^{1/3}\Gamma_{1}^{8/3}\ {\rm erg.s}^{-1},
\end{equation}
and 
\begin{equation}
T_{\rm obs}={5\over 1+z}\,\epsilon_{\rm th}^{1/4}\left(1+\sigma\right)^{2/3} R_{0,7}^{1/6}{\dot E}_{50}^{-5/12}\Gamma_{1}^{8/3}\ {\rm keV}.
\end{equation}
It is illustrative to consider the ratio $L_{\rm ph}/T_{\rm obs}$, since it is independent of $\Gamma$ and relatively insensitive to the other model parameters, yet can still be compared directly to observations. We find
\begin{equation}
L_{\rm ph}/T_{\rm obs}=2.8\,10^{47}(1+z)\,R_{0,7}^{1/2}\,(\epsilon_{\rm th}{\dot E}_{50})^{3/4}\ {\rm erg.s^{-1}.keV^{-1}}\ .
\end{equation}
Applying this expression to GRB 070110 during the plateau phase with $L_{\rm ph}=10^{48}$ erg.s$^{-1}$, $T_{\rm obs}\approx 1$ keV and $R_{0}=10^7$cm  we obtain
$\epsilon_{\rm th}{\dot E}_{50}\approx 1$. 
With a plateau duration of about 5800 s (rest frame), 
the total (isotropic equivalent) energy released by the central engine during this phase should be $E=6\,10^{53}/\epsilon_{\rm th}$ erg, which,
for $\epsilon_{\rm th}$ smaller than unity, becomes uncomfortably large.
Notice that we have assumed here that the emitted power during the plateau falls predominantly in the [0.1-10] keV range observed by {\it Swift} XRT. If this is not the case the energetic requirements become even larger. The possibility of powering the observed plateau from a non-dissipative photosphere is therefore disfavoured. In what follows we thus turn to a discussion of dissipative photospheres.
\subsection{Flow propagation with dissipation}
\label{sec:dissipation}
A larger efficiency can be achieved if dissipation takes place in the outflow between the initial 
injection radius $R_0$ and the photosphere, $R_{\rm ph}$. Furthermore, if the jet undergoes strong dissipation below the photosphere,
it could result in a strong photospheric signal, even if the flow is initially strongly magnetically dominated. Hence, the required photospheric component in our model does not necessarily impose a baryonic jet. The general tendency is that the lower the dissipation occurs below the photosphere (i.e. the higher the optical depth at the dissipation radius) and the lower the initial energy per Baryon, the more thermal-like would be the resulting signal. This has been studied in many works, such as \citep{GianniosSpruit2005,Giannios2006,Giannios2007,Beloborodov2011,Vurm2011}.
If dissipation happens mostly at a large optical depth, 
it can increase the efficiency but the resulting spectrum remains close to a blackbody. However, if dissipation extends all the way 
to the photosphere, Comptonization of the thermal photons by energetic electrons can replace the exponential cut-off of the Planck function
by an extended power-law.

In the first case the expected spectrum will be a combination of a blackbody component from the plateau with a power-law component from the
underlying afterglow. This possibility, is however unlikely, given that attempting to fit the XRT spectrum at the time of the plateau with a blackbody + power law model, we find that in the spectral interval 0.5-2keV (observer frame), the thermal flux is at most (at the 3 sigma level) 0.28 of the flux of the power law in the same band. This clearly demonstrates that the plateau emission cannot be dominated by a pure blackbody component.

The observed X-ray spectrum at the time of the plateau, after correcting for dust absorption, was fitted with a simple power law $F_{\nu} \propto \nu^{-1}$ \citep{Troja2007}. Note that since the burst is high-redshift ($z=2.35$), the dust absorption is dominated by the intergalactic rather than the galactic component. There is therefore relatively small uncertainty in the dust reduction process. 
However, the $\nu^{-1}$ power law cannot be the end of the story. This is because
its extension to the optical band, would strongly over-predict the observed flux in that band.
In addition, as mentioned in \S \ref{sec:magnetar}, \S \ref{sec:nodiss} the energetic requirements for the plateau are already very large. Unless a cut-off or a significant break
in the spectrum both below and above the XRT range are invoked, these requirements would quickly become unmanageable.
This suggests that even if the plateau emission is not purely thermal, it clearly cannot be a pure power-law as well,
and it is useful to consider other possibilities.
Indeed we find that a Planck-like function where the exponential cut-off is replaced by a power-law approaching $F_{\nu}\propto \nu^{-1}$ (which can be naturally obtained assuming strong Comptonization effects take place close to the photosphere, see \citealt{Giannios2007}) overcomes the difficulties described above and is consistent with the data, so long as the temperature is $\lesssim 0.5$ keV.

In the case of a Comptonized photosphere we introduce two dimensionless parameters to respectively account for energy dissipation and deviation from a blackbody spectrum.
These are respectively, $\epsilon_{\rm rad}$ and $\lambda$, which are defined by
\begin{equation}
\label{eq:Lth}
L_{\rm th}=\epsilon_{\rm rad}{\dot E}=4\pi\left({R_{\rm ph}\over \Gamma}\right)^2\sigma_{\rm S}\left({\lambda E_{\rm p}\over 4 k}\right)^4\ ,
\end{equation}
where $\sigma_{\rm S}$ and $k$ are the Stefan and Boltzmann constants and $E_{\rm p}$ is the peak energy of the spectrum in the source rest frame. The parameter $\lambda$
is smaller or equal to unity, with equality corresponding to a pure (not Comptonized) Planck spectrum;
$\epsilon_{\rm{rad}}$ is equal or larger than the efficiency of a non-dissipative photosphere, which using Eq. \ref{eq:Lph2} gives
\begin{equation}
\epsilon_{\rm{rad}} \geq 6.5\,10^{-2}\,\epsilon_{\rm th}\left(1+\sigma\right)^{2/3}R_{0,7}^{2/3}{\dot E}_{50}^{-2/3}\Gamma_{2}^{8/3} .
\end{equation}
From Eq. \ref{eq:Lth} and the expression of the photospheric radius
\begin{equation}
\label{eq:Rph2}
R_{\rm ph}\approx {\sigma_{\rm T}\dot E \over 8\pi m_{\rm p} c^3 \Gamma^3(1+\sigma)}\ ,
\end{equation}
we can obtain the Lorentz factor at the photosphere 
\begin{eqnarray}
\label{eq:GAph}
\Gamma&=&20\,{\lambda^{1/2}\over \epsilon_{\rm rad,-2}^{1/4}}{\left( 1+\sigma\right)}^{-1/4}L_{\rm th,48}^{1/8}E_{\rm p,obs}^{1/2}\\
&=&20\,{\lambda^{1/2}\over \epsilon_{\rm rad,-2}^{1/8}}{\left(1+\sigma\right)}^{-1/4}{\dot E}_{50}^{1/8}E_{\rm p,obs}^{1/2}\,.
\end{eqnarray} 
The Lorentz factor of the slower moving material is therefore rather insensitive to the model parameters. As mentioned in \S \ref{sec:nodiss}, if the flow has not reached the saturation radius at the photosphere, terminal Lorentz factor could be larger (and Eq. \ref{eq:GAph} would be considered as a minimum on its actual value). However, this possibility is unlikely for $\Gamma_s\lesssim 20$ and would in any case lead to a peak which is significantly higher than the observed band. We therefore do not consider this possibility in the following.
Plugging Eq. \ref{eq:GAph} back into Eq.\ref{eq:Rph2} we find
\begin{equation}
R_{\rm ph}=7.5\,10^{12}{\epsilon_{\rm rad,-2}^{3/8}\over \lambda^{3/2}}{\left(1+\sigma\right)}^{-1/4}{{\dot E}_{50}^{5/8}\over E_{\rm p,obs}^{3/2}}\ {\rm cm},
\end{equation}
with a corresponding geometric time-scale  
\begin{equation}
t_{\rm geo}^{\rm obs}={R_{\rm ph}(1+z)\over 2\,c\Gamma^2}={\epsilon_{\rm rad,-2}^{5/8}\over \lambda^{5/2}}{\left( 1+\sigma\right)}^{1/4}{{\dot E}_{50}^{3/8}\over E_{\rm p,obs}^{5/2}}\ {\rm s}\ ,
\end{equation}
where the observed peak energy, $E_{\rm p,obs}$, is in keV (a redshift $z=2.35$ was assumed). The actual values of $\Gamma$, $R_{\rm ph}$ and $t_{\rm geo}$ depend on the uncertain parameters
$\epsilon_{\rm rad}$, $\lambda$ and $(1+\sigma)$. Considering for example a Comptonized photosphere with $\lambda\sim 0.3$ and a magnetization parameter $(1+\sigma)\sim 2$, the Lorentz factor is only reduced by $\approx 2$ compared to the case with $\lambda=(1+\sigma)=1$, while the photospheric radius is increased by $\approx 5$ and the geometrical time-scale becomes $\approx 24$ times smaller.     

It can be seen that, except for very small values of $\lambda$ or/and $E_{\rm p,obs}$, the geometric time scale remains much shorter than the observed decay
time at the end of the plateau, which is close to 1.5 hours. This indicates that the observed decline effectively corresponds to the physical extinction 
of the central engine.

\section{Revere and forward shock contributions}
\label{sec:RevFor}
\subsection{Dissipated power and reverse shock re-brightening}
The relativistic ejecta emitted by the central engine of GRB 070110 is structured in two parts. The first one, with a Lorentz factor of a few hundreds is decelerated first and is responsible for the standard afterglow (i.e. excluding the internal plateau) until the second slower part with 
a Lorentz factor of a few tens is able to catch up. This second part may be structured, with initially an outflow lasting
1.5 h (rest frame) with a nearly constant $\Gamma$ to make the plateau, followed by a tail where $\Gamma$ decreases to non relativistic values close to unity. 
This slow part will add energy to the forward shock and, if its magnetization is low enough, it will be crossed by a reverse shock. The crossing time 
of the region with a constant Lorentz factor will be short, leading to a sharp increase in dissipated power, which may explain the bump in the XRT light curve that follows the plateau.

\begin{figure}
	\includegraphics[width=0.45\textwidth]{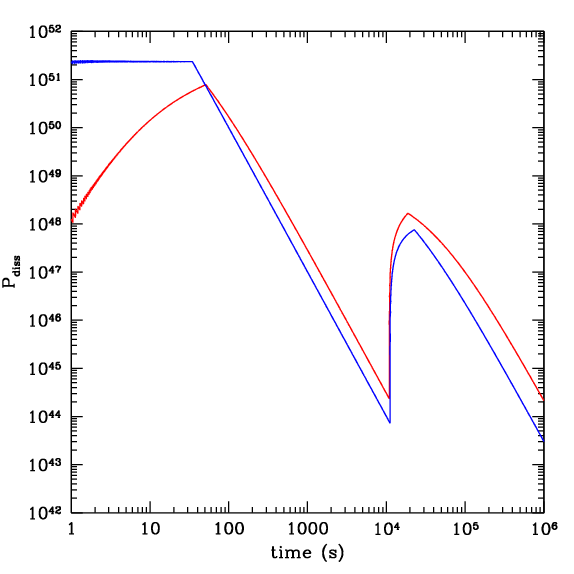}
	\caption{Dissipated power as a function of time for a fast shell of material ($\Gamma\approx 200$) that decelerates into the external medium and is eventually overtaken by a slower outflow, with $\Gamma=20$.
		Results are shown for a uniform medium of density $n=30$ cm$^{-3}$ (red) and a stellar wind with $A_*=2.4$ (blue).}
	\label{fig:disspower}
\end{figure}

To test this possibility we have calculated the dissipated power in the reverse shock following the method described by \cite{Genet2007}. We 
adopt a Lorentz factor $\Gamma_f=200$ and an injected kinetic power $\dot E_f=2\times 10^{52}$ erg.s$^{-1}$ 
for a duration of 30 s (rest frame) in the ejecta responsible for the prompt emission. The Lorentz factor can be expected to
be highly variable in this initial part of the flow and the value $\Gamma_f=200$ therefore represents some 
typical average. This fast part is then
followed by a slower long lasting outflow with $\Gamma_s=20$, $\dot E_s=2\times 10^{49}$ erg.s$^{-1}$ and a duration of 5800 s, which makes the 
plateau.  
The power dissipated when the reverse shock crosses this material is shown in Fig.\ref{fig:disspower} for a uniform medium of 
density $n=30$ cm$^{-3}$ and a stellar wind with $A_*=2.4$. It is proportional to the bolometric luminosity
of the reverse shock in case the radiative efficiency is constant and the electrons are in the ``fast cooling'' regime.

The sharp increase in dissipated power happens when the reverse shock encounters the low $\Gamma$ tail of the ejecta.
The Lorentz factor of the material in the reverse shock is approximated by
\begin{equation}
\Gamma_r(t_b)=\sqrt{\Gamma_f(t_b)\Gamma_s \frac{M_s \Gamma_s+(M_f+\Gamma_i(t_b) M_{ex}(t_b))\Gamma_f(t_b)}{M_s \Gamma_f(t_b)+(M_f+\Gamma_i(t_b) M_{ex}(t_b))\Gamma_s}}
\end{equation}
where $M_s=E_s/\Gamma_sc^2$, $M_f=E_f/\Gamma_f(t=0)c^2$ are the masses of the slow and fast shells, $M_{ex}$ is the mass of external medium swept up by the shock, $\Gamma_i\approx \Gamma_f$ is the Lorentz factor associated with internal motions in the shocked external medium (see \cite{Genet2007} for a detailed description) and $t_b$ is the time of the bump, i.e. the time of collision between the slow and fast shells.
To estimate this time, we first relate the radius and (source frame) time for the fast moving shell 
\begin{equation}
\label{eq:tf}
t=\int^R \frac{dR}{2c \Gamma^2}=\int^{R_d}\frac{dR}{2c \Gamma_f^2}+\int_{R_d}^R\frac{R^{2\epsilon}dR}{2c \Gamma_f^2 R_d^{2\epsilon}}=\frac{t_d}{2\epsilon+1}\bigg( \bigg(\frac{R}{R_d}\bigg)^{2\epsilon+1}+2\epsilon\bigg)
\end{equation}
where the subscript $d$ denotes quantities at the deceleration radius and $\epsilon=3/2$ for ISM ($\epsilon=1/2$ for wind). The slow moving jet is moving in the path evacuated by the fast shell, and thus propagates with a constant Lorentz factor: $R=2ct_s\Gamma_s^2$. Requiring that the two shells reach a shared radius at the same time, we plug the latter expression for $R$ in Eq. \ref{eq:tf} and obtain an equation for the collision time, $t$
\begin{equation}
t=\frac{t_d}{2\epsilon+1}\bigg( 2 \epsilon + \bigg(\frac{\Gamma_s^2t}{\Gamma_f^2 t_d}\bigg)^{2\epsilon+1}\bigg).
\end{equation}
At the limit of $t \gg t_d$ we obtain 
\begin{equation}
t=t_d (2\epsilon+1)^{1\over 2 \epsilon} \bigg(\frac{\Gamma_f}{\Gamma_s}\bigg)^{2+\frac{1}{\epsilon}}
\end{equation} which can be re-written as
\begin{eqnarray}
\label{eq:bump}
&t_b&=\left(3 E_{f,\rm K}\over 8\pi n m_{\rm H} c^5 \Gamma_s^8\right)^{1/3}
=1.4 \times 10^5 \left({E_{f,53}\over n}\right)^{1/3} \Gamma_{s,1}^{-8/3}\ \ {\rm s}
\\
&t_b&=\left(E_{f,\rm K}\over 4 \pi A c^3 \Gamma_s^4\right)
=6 \times 10^4 \left({E_{f,53}\over A_*}\right) \Gamma_{s,1}^{-4}\ \ {\rm s}
\end{eqnarray}
for a uniform medium and a stellar wind respectively, where
the index $s$ denotes the slower moving material and $f$ the fast moving material.
As shown by \cite{KumarPiran2000}, this is the time necessary for the fast moving material to decelerate to $\Gamma_s/2$ for a uniform medium ($\Gamma_s/\sqrt{2}$ for wind). 
Its value strongly depends on $\Gamma_s$ but for the parameters of GRB 070110 required to explain the plateau via photospheric emission 
($\Gamma_s \sim 20$, typically; see \S \ref{sec:photo}) we can expect a bump at $\sim 10^{4}$ s in the burst rest frame (as indeed implied by observations) 
provided that the density of the external medium is sufficiently large , i.e. $n\approx 30$ cm$^{-3}$ or $A_*\approx 2.4$ for a wind).
Due to the strong dependence of $t_b$ on $\Gamma$, even a slight change in $\Gamma$ would have a large effect. For instance, if the terminal Lorentz factor was 25 instead of 20, 
$A_*$ and $n$ could be reduced by a factor of 2, to about $1$ and $16$ cm$^{-3}$ respectively.

\subsection{X-ray and visible light curves}
In order to calculate the light-curves at a given observed frequency, one must specify the microphysical shock parameters $\epsilon_e,\epsilon_B$ as well as the power law slope for the electrons' energy distribution, $p$.
Since the forward shock is an ultra-relativistic shock, whereas the reverse shock is only mildly relativistic, it is quite likely that these parameters vary from one environment to the other. Many afterglow studies have constrained the values of the microphysical parameters at the forward shock $\epsilon_{e,f},\epsilon_{B,f}$. $\epsilon_{e,f}$ is strongly constrained by both afterglow modelling \citep{Santana2014} and numerical simulations of shock acceleration \citep{Sironi2011} to $\epsilon_{e,f}\approx 0.1$. Furthermore, clustering of LAT light-curves implies that the distribution around this value is quite narrow \citep{Nava2014} and finally it cannot be much lower in order to avoid un-physically large energetic requirements for the blast wave.
The value of $\epsilon_{B,f}$ is less certain. Various recent studies suggest $10^{-6}<\epsilon_{B,f}<10^{-2}$ \citep{Lemoine2013,Wang2013,BD2014,Santana2014,Zhang2015,
	Beniamini2015,Beniamini2016,Burgess2016}.
For the reverse shock, there is much less data and the situation is less clear. However following the available estimates in the literature \citep{Fan2002,Kobayashi2003,McMahon2006,Shao2005} we adopt here as canonical values $\epsilon_{e,r}=0.1$ and leave $\epsilon_{B,r}$ as a free parameter.
Reducing either $\epsilon_{e,r}$ or $\epsilon_{B,r}$ would decrease the strength of the bump due to the reverse shock as will be shown explicitly below (Eq. \ref{eq:aboveC}).
Finally, we take $p=2.15$ for both the reverse and forward shock.
This value of $p$ is typical for GRB modelling and is well constrained by comparing the X-ray and optical fluxes during the regular forward afterglow stage (at times later than the plateau and the following bump) and assuming that both are above the cooling break. This assumption is further supported by the fact that the flux in the two bands decline at the same rate as a function of time \footnote{The alternative possibility, that both bands are below the cooling break would require rather irregular values of $p>3$ and smaller values of $\epsilon_B$ for the emitting material.}. The afterglow of GRB 070110 does not follow the regular closure relations expected in the forward afterglow scenario (see \S \ref{sec:obs}). In other words, the value of $p$ implied by the spectrum does not match the value implied by the temporal behaviour. The most natural way of resolving this apparent contradiction, as indeed pointed out by \cite{Troja2007}, is to invoke late-time continuous energy injection. This would both enhance the forward shock emission as well as lead to a prolonged reverse shock emission, that under certain conditions could dominate the post-plateau light-curve. 
In fact this can be viewed as further evidence for the association of the bump itself with energy injection (and thus of the plateau with emission from this extra material). In order to match the observed post-bump decline, we adopted here a tail component to the slow material, lasting for 400 s (source frame), with $\dot{E}_t=\dot{E}_s$, and with $\Gamma$ declining from $\Gamma\approx 20$ down to $\Gamma\approx 1$ with $\Gamma(t_{inj})\propto (t_f-t_{inj})^{2/11}$ where $t_{inj}$ is the time of injection of a given shell of matter and $t_f$ is the time of injection for the final shell.

Given these parameters, we calculate the synchrotron flux from the forward shock \citep{Granot2002}. For the reverse shock we estimate the physical conditions as specified in \cite{Genet2007}.
In both cases we include possible IC losses including Klein -Nishina corrections (see \citealt{SariEsin2001,Nakar2009,Beniamini2015} for details). IC losses become significant for small values of $\epsilon_B$.
Importantly, for these low magnetizations, the X-ray emission from the forward shock is very inefficient due to IC losses of the electrons emitting in the fast cooling synchrotron regime \citep{SariEsin2001}.
This is in fact essential in our model in order to allow for a significant contribution from the reverse shock at the time of the bump (and after), without investing an amount of energy in the slow shell that would be considerably larger than the energy stored in the fast component. Indeed, the jet approximately dissipates (all of) its energy on a time-scale of order the dynamical time, so {\it if} this energy is efficiently radiated as synchrotron in the X-ray regime, the only way to significantly increase the flux would be to significantly increase the energy.
Avoiding an energy that is too large in the slow shell, is important both on grounds of the total available energy budget and more directly, since such a large energy is not seen during the plateau emission.
The low values of $\epsilon_B$, motivated by values found in the literature, suppress the synchrotron flux due to IC losses, and thus can allow for a significant contribution from the reverse shock without investing huge amounts of energy in the latter component (assuming that in the reverse shock, $\epsilon_B$ is not so small as in the forward shock).

\begin{figure*}
	\begin{center}
		\begin{tabular}{cc}
			\includegraphics[width=1.0\textwidth]{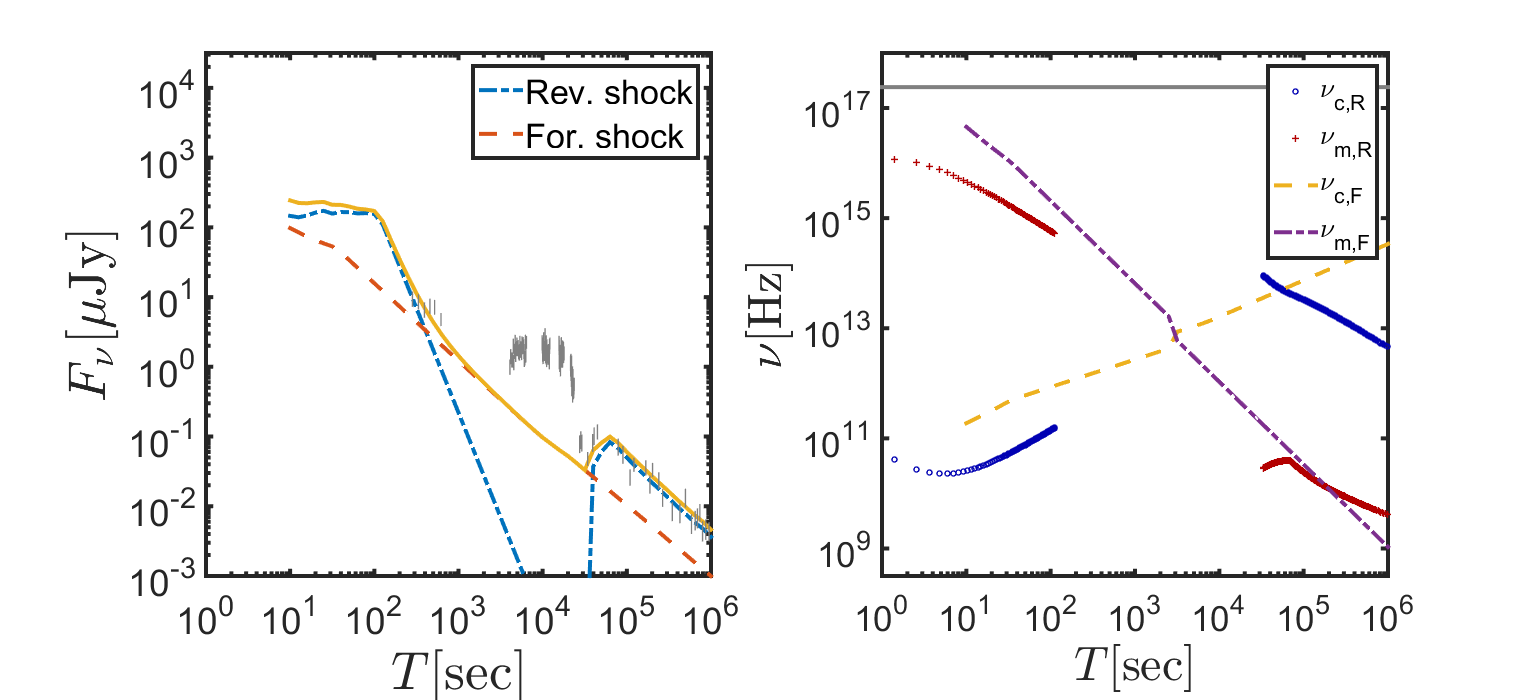}\\
			\includegraphics[width=1.0\textwidth]{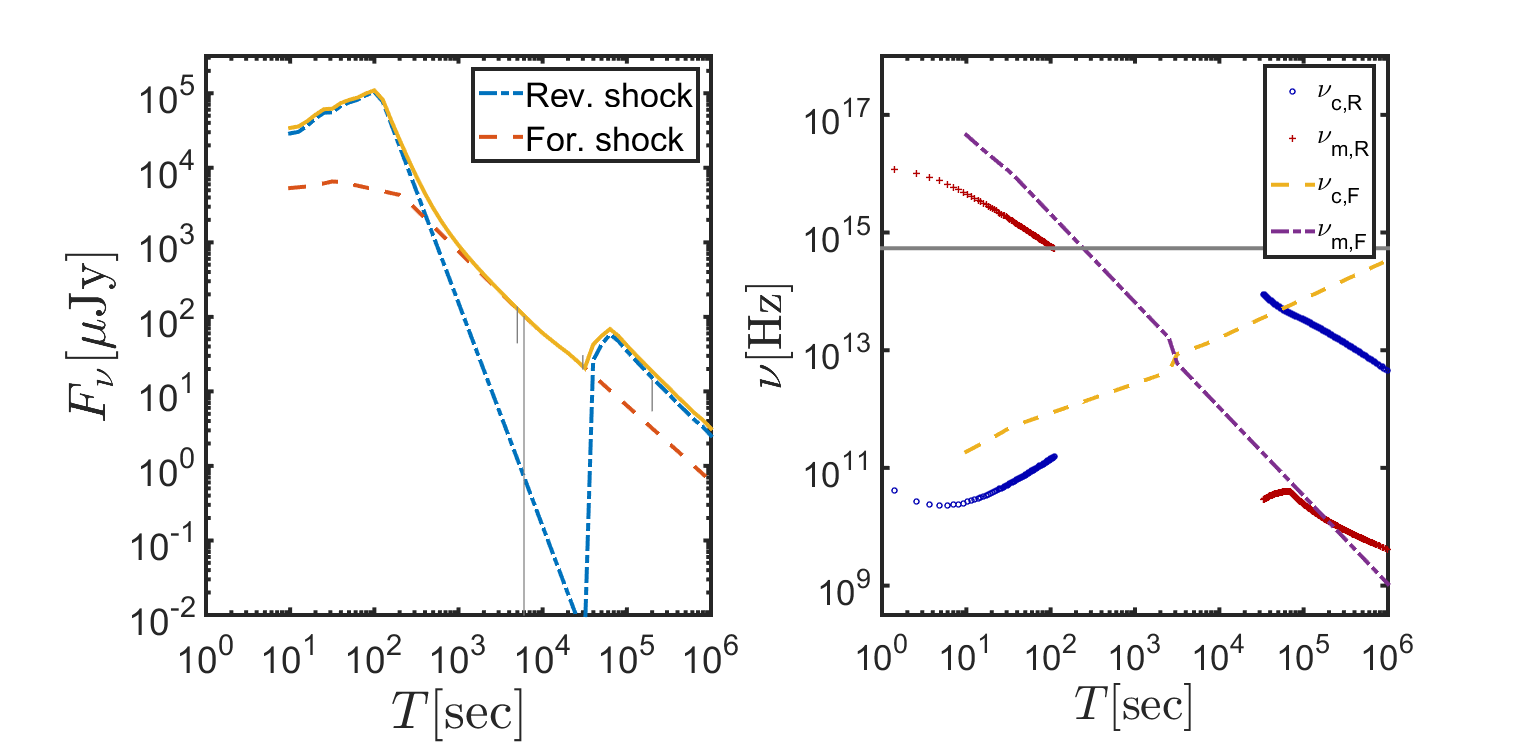}
		\end{tabular}
	\end{center}
	\caption{Light-curves corresponding to reverse and forward shock emission assuming the ejecta responsible for the prompt emission is followed by slower moving outflow with $\Gamma_s=20$, $\dot E_s=2\times 10^{49}$ erg.s$^{-1}$ and lasting 6000 s, which makes the plateau. The tail of this material is characterized by the same $\dot{E}$, lasting 400 s and with a declining 
		Lorentz factor from $\Gamma\approx 20$ down to $\Gamma\approx 1$ with $\Gamma(t_{inj})\propto (t_f-t_{inj})^{2/11}$
		The reverse shock crossing this material accounts for a bump in the afterglow data. The time of the bump, as well as the flux levels during and after the bump match the observed levels.
		We assume here $A_*=2.2,\epsilon_{e,f}=0.1,\epsilon_{B,f}=2.5\times 10^{-5},\epsilon_{e,r}=0.1,\epsilon_{B,r}=4\times 10^{-3},p=2.15$. Top panels correspond to the 1keV light-curve and bottom panels to the 1eV light-curve.
		{\bf Left:} Light-curve for the reverse shock (blue dot-dashed line) and forward shock (red dashed line). A yellow solid line depicts the sum of the reverse and forward shock contributions.{\bf Right:} Evolution of $\nu_{c,R}$ (blue circles), $\nu_{m,R}$ (red pluses), $\nu_{c,f}$ (yellow dashed line) and $\nu_{m,f}$ (purple dot-dashed line). The observed frequency is depicted by a horizontal line.}
	\label{fig:shocklightcurve}
\end{figure*}

\begin{figure*}
	\begin{center}
		\begin{tabular}{cc}
			\includegraphics[width=1.0\textwidth]{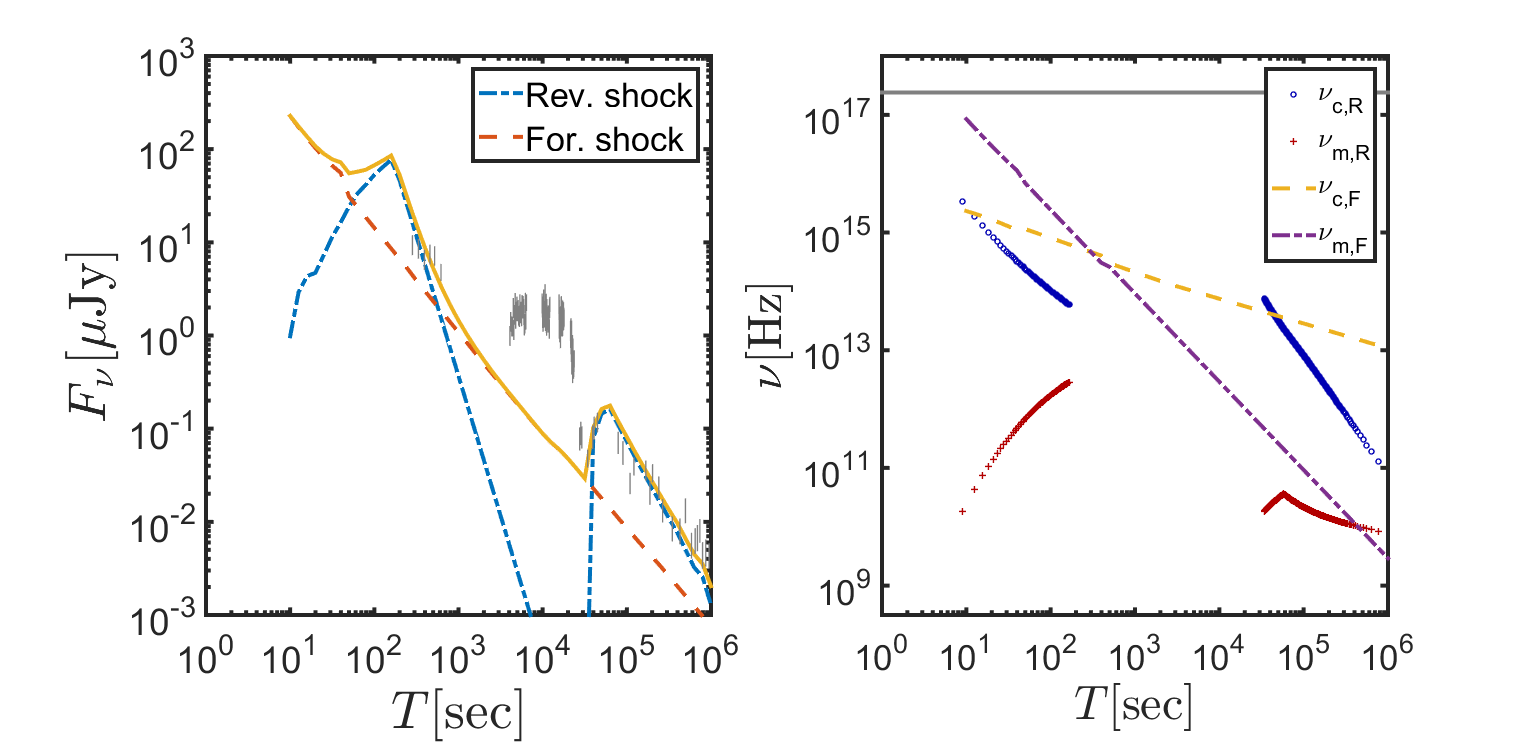}\\
			\includegraphics[width=1.0\textwidth]{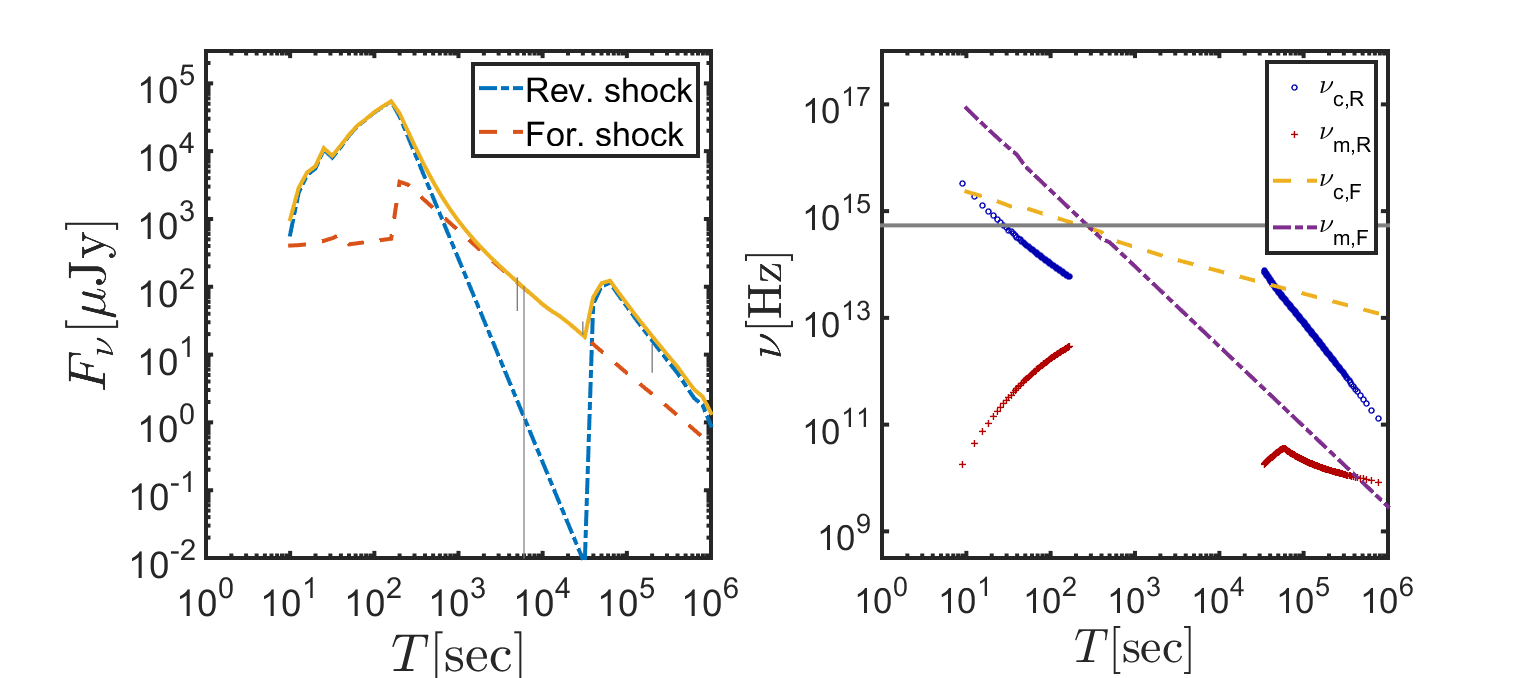}
		\end{tabular}
	\end{center}
	\caption{Same as Fig. \ref{fig:shocklightcurve} for an ISM medium with $n_0=30\mbox{ cm}^{-3}$. The choice of burst parameters is the same as in Fig. \ref{fig:shocklightcurve} except for $\epsilon_{B,f}=10^{-4},\epsilon_{B,r}= 2\times 10^{-2}$.}
	\label{fig:shocklightcurveISM}
\end{figure*}

Using the values of the parameters described above, as well as $\epsilon_{B,f}=2.5\times 10^{-5},\epsilon_{B,r}=4\times 10^{-3}$ for the wind case ($\epsilon_{B,f}=10^{-4},\epsilon_{B,r}= 2\times 10^{-2}$ for ISM), we compute the forward and reverse shock light-curves in the optical and X-ray bands.
The results for the canonically assumed parameters are shown in Fig. \ref{fig:shocklightcurve} for a wind medium and in Fig. \ref{fig:shocklightcurveISM} for an ISM environment.
For both a wind or ISM environment, a bump due to the reverse shock contribution can be seen in the X-ray and optical bands. Both the time and magnitude of the bump match the observed data.
At $t\gtrsim 10^5$sec the flux is dominated by the reverse shock emission from the tail of material trailing behind the matter contributing to the plateau + bump.
Since at both the optical and X-ray bands, the reverse shock contribution is expected to be in the fast cooling regime, one does not expect a change in the spectral slope during the bump, which indeed matches the available observations at those times.

To understand the general dependence of the bump appearance on the model parameters, it is useful to consider simplified expressions for the reverse and forward shock luminosities.
For typical choices of the parameters, the X-ray and optical emitting 
electrons in both the reverse and forward shock are fast cooling (see Fig. \ref{fig:shocklightcurve}) and $\epsilon_{B,f} \ll \epsilon_{e,f}$ (see discussion above). The ratio of reverse to forward shock flux at the time of the bump can then be approximated by:
\begin{equation} 
\label{eq:aboveC}
\frac{F_r}{F_f}\!=\!
\left\{ \!
\begin{array}{l}
21f(p)n_{0,1.5}^{6-p \over 12} E_{f,53.5}^{p-20 \over 20}\epsilon_{e,f,-1}^{2-p}\epsilon_{B,f,-4}^{-(2+p)\over 4} \dot{E}_{s,50}^{32+p \over 32}
t_{inj,4}^{32-7p\over 32}\\ \Gamma_{s,1}^{9-5p\over 3} \epsilon_{e,r,-1}^{p-1} \epsilon_{B,r,-2}^{p\over 4} \quad \mbox{for ISM}\\
\\
8\tilde{f}(p)A_{*,1}^{6-p \over 4} E_{f,53.5}^{7p-32 \over 16}\epsilon_{e,f,-1}^{2-p}\epsilon_{B,f,-4}^{-(2+p)\over 4} \dot{E}_{s,50}^{16-3p \over 16}
t_{inj,4}^{1\over 8}\\ \Gamma_{s,1}^{5-2p} \epsilon_{e,r,-1}^{p-1} \epsilon_{B,r,-2}^{p\over 4} \quad \mbox{for wind}\\
\end{array} \right.
\end{equation}
where $f(p),\tilde{f}(p)$ are numerical factors of order unity, depending weakly on $p$. Naturally, a large energy in the slow moving material as compared with that in the fast moving material as well as a higher density of the external medium lead to an increase in this ratio. Furthermore, as noted above, a small value of $\epsilon_{B,f}$ suppresses the synchrotron flux from the forward shock and so also increases this ratio.

\section{Discussion and conclusion}
\label{sec:summary}
GRB 070110 was a peculiar event. While its prompt phase was not remarkable, showing a few pulses and ending with the usual 
early steep decay of temporal index close to $-3$, it was followed by a very flat plateau interrupted by an extremely 
steep drop with the X-ray flux declining as $t^{-9}$. One (or two) bump(s) were observed just after this drop with the flux 
increasing by a factor 2-3 before a more regular power-law decline of index $-0.6$ eventually followed until the end of XRT observations after three weeks.

We interpreted the initial early steep decay as the high latitude emission from the last flashing shell of the prompt phase. This implies a large 
emission radius for the prompt phase $R\sim 2\,c\Gamma^2 \tau$ where $\tau$ is the duration of the prompt activity and therefore favours dissipation from internal shocks or
magnetic reconnection above the photosphere. For the plateau we first re-considered the possibility that it could be powered by a magnetar but 
concluded that it was not likely due to tight constraints on the energy budget and a fine-tuning requiring the collapse to happen at a time just below
the spin-down time of the magnetar.

We instead favour a scenario where the plateau results from the photospheric emission of a continuous outflow of moderate Lorentz factor emitted from a central black hole 
(probably powered by accretion, the prompt phase being powered by the Blandford-Znajek mechanism). The geometric timescale at the photospheric radius is 
then so short that the very steep decay at the end of the plateau is unlikely to be dominated by high-latitude emission from the photosphere, and rather, should correspond to the actual shut-down of the central source activity. Some dissipation must
take place in the flow, both to increase the efficiency compared to a pure adiabatic evolution and to transform the original black-body spectrum by adding a power-law tail at high energy.

The propagation of the reverse shock in the material producing the plateau generates a bump in the light curve since, due to the assumed nearly uniform Lorentz factor, 
it is crossed by the shock in a short time scale. If, in addition, the outflow ends with a tail where the Lorentz factor decreases to values close to
unity in a few hundreds of seconds, the reverse shock contribution will be long-lived and could account for the late shallow temporal slope in both the X-rays and visible.                    

If this global picture is correct it leads to a certain number of consequences: 
\begin{itemize}
	\item
	The decay at the end of a period of activity can be used to constrain the emission radius and therefore the dissipative process and radiation mechanism.
	In the case of GRB 070110, it predicts that the prompt phase resulted from dissipation above the photosphere (from internal shocks or magnetic reconnection) while the plateau phase 
	was of photospheric origin. Photospheric emission during the prompt phase was therefore sub-dominant, for example because the ejecta was still magnetically dominated (see also \citealt{Hascoet2013}). 
	Conversely, during the plateau, dissipation was mostly confined below the photosphere. This could be simply due to the increase of the photospheric radius at low 
	Lorentz factor (even if the injected power decreases). For example, in the case of shock dissipation, the ratio $R_{\rm IS}/R_{\rm ph}\propto \Gamma^5$ could easily fall below unity 
	at low $\Gamma$.      
	\item
	If the bump following the plateau indeed results from the propagation of the reverse shock in the low $\Gamma$ part of the ejecta, it shows that the magnetization was low ($\sigma \lesssim 0.1$) at that time,
	either because it was originally low or has decreased during the propagation of the flow. 
	\item 
	The peculiarities of GRB 070110 (internal plateau, very steep drop at the end of the plateau, shallow decline at late times) are unusual. This could
	indicate that ({\it i}) a prolonged activity of the central engine is rare, or ({\it ii}) that plateaus, even if present, radiate mostly in UV because the Lorentz factor of the  
	long-lasting outflow is lower than 10-20 or are dimmer due to less injected power or to
	adiabatic cooling without additional dissipation in the outflow.
	\item If other bursts with similar internal plateaus are found, we might be able to detect bumps and/or deviation from closure relations in their afterglow after the plateau. At 
	present, only a few events with flat plateaus followed by a steep decline have been found (such as GRB 060413 or GRB 120213A) but the decay after the plateau is not as 
	steep as in GRB 070110 (with a temporal index of $-2.5$ to $-3$), which does not necessarily require a photospheric origin.
	\item More speculatively, one may postulate that the high density environment apparently implied by the bump and late-time afterglow, is physically related to the prolonged activity of the central engine. This may happen, for instance, if due to a more violent mass ejection towards the end of the star's life it ejected more matter that ended up being accreted to the newly formed black hole. This possibility may be quantitatively testable with numerical simulations or observationally if future GRBs reveal similar properties to those of GRB 070110.
\end{itemize}
The study of GRB 070110 illustrates how bursts with non-standard features can help decipher the dissipation processes and radiation processes
at work in GRBs. Any indication of the presence of shocks constrains the magnetization. The variability timescales puts limits on the angular spreading and therefore 
on the emission radius. Finally, plateaus can be related to late source activity. One may hope that the diversity in early afterglow behaviours, which can be viewed as a source of confusion, 
may also provide tools for a better understanding of GRB physics. 

\begin{acknowledgements}
     We thank Johannes Buchner for applying the spectral fits to the plateau data using the Comptonized photospheric model as described in \S \ref{sec:dissipation}. We also thank the anonymous referee for their detailed comments and suggestions on the manuscript. PB is supported by a Chateaubriand Fellowship.
\end{acknowledgements}

\end{document}